\documentclass[runningheads]{svmult}

\usepackage{makeidx}   
\usepackage{graphicx}  
\usepackage{subeqnar}  
\usepackage{multicol}  
\usepackage{physprbb}  
\makeindex             

\begin{document}
\title*{UVES detailed chemical abundances in the Sgr dSph and the CMa overdensity}
\toctitle{UVES detailed chemical abundances
\protect\newline in the Sgr dSph and the CMa overdensity }
%
%
\titlerunning{Uves abundances in Sgr and CMa}
%
\author{Luca Sbordone\inst{1,2}
}
\authorrunning{L. Sbordone et al.}
%
%
\institute{ESO -- European Southern Observatory 
\and
Universit\'a di Roma 2 ``Tor Vergata'' 
}

\maketitle              


\section{Introduction}
In this contribution, we present detailed chemical abundances on two stellar systems presently believed to be undergoing tidal merging with the Milky Way (MW). The first one is the Sagittarius Dwarf Spheroidal (Sgr dSph, \cite{sbo:ibata95}), a massive  ($10^8$ M$_{\odot}$) dSph orbiting along a very short period ($<~1 GYr$) almost polar orbit inside the Halo, along which is slowly dissolving in a huge stellar stream \cite{sbo:majewski03}.
The second object is the recently discovered, and still controversial CMa dwarf galaxy \cite{sbo:martin03}, a heavily degraded overdensity embedded in the MW thick disk, believed to be the residual of an in-plane accretion of an object of mass comparable to the Sgr dSph. This interpretation is still controversial, since \cite{sbo:momany04} claimed that the structure almost disappears, when disk warping and/or flaring is properly taken into account in the modeling of the MW contribution.

\section{Data and analysis}
For the Sgr dSph we present the UVES DIC1 spectra for 12 giants. Complete analysis of two of them has already been published \cite{sbo:bonifacio00}, while for the other ten only iron and $\alpha$-elements abundances have been published so far (see \cite{sbo:bonifacio04}). Details on the reduction and analysis procedures, and physical parameters for the stars are provided in \cite{sbo:bonifacio04}, but they can be briefly resumed here: the spectra have been analyzed by means of LTE, one dimensional atmosphere models, using ATLAS, WIDTH and SYNTHE codes (see \cite{sbo:kurucz93} and \cite{sbo:sbordone04}). T$_{eff}$ for the stars are in the range 4800 - 5050 K, $\log$ g between 2.3 and 2.7. We analyzed abundances of proton capture (Na, Al, Sc, V), $\alpha$ (Mg, Si, Ca, Ti), Iron-peak (Cr, Fe, Co, Ni, Zn) and heavy neutron-capture (Y, La, Ce, Nd) elements.

Immediately after the discovery of the CMA overdensity we obtained DDT time at VLT -- FLAMES for an exploratory study of the CMa population underlying the galactic open  cluster NGC 2477, identified by \cite{sbo:bellazzini04}. Here we present the results relative to the investigation of the 7 UVES spectra, the results relative to the stars observed through the MEDUSA fibers are presented in \cite{sbo:zaggia04}. Four of the UVES stars had radial velocities or absolute magnitudes incompatible with the CMa overdensity; a paper detailing the analysis and results of the remaining three objects has been submitted (see \cite{sbo:sbordone04b}). The analysis has followed the same method above described for Sgr dSph stars. Two stars appear to be giants (log g 2.3 and 2.8, T$_{eff}$ 4990 and 4994 K), while the third one is a subgiant (T$_{eff}$ 5367, log g 3.5).  For these stars, in addition to the elements above listed,  we also analyzed Co, Cu, Ba and Eu.

\section{Results}

The most striking feature observed in Sgr dSph is the presence of a metal rich ([Fe/H] between -1 and 0) and young population, showing the $\alpha$-elements underabundance now recognized as typical of dSph (see \cite{sbo:venn04}). Besides that, we observe a strong (up to 0.8 dex) overabundance of n-capture elements (La, Ce, Nd) with respect to iron, a significant Na and Al underabundance, and an intriguing Ni deficiency. All these chemical features appear to be typical of Sgr dSph and of its associated systems like Terzan 7 and Pal 12 (e.g. \cite{sbo:cohen04}). Moreover, we observed an important underabundance of Zn ([Zn/Fe] $\sim$ -0.4). This may undermine the parallelism between dSph and the DLAs, where Zn is used as a proxy for Fe.

Of the three stars in the CMa overdensity, the subgiant appears the most interesting one. First of all, as can be seen in \cite{sbo:zaggia04}, contamination of the sample is a significant issue here, and only the subgiant, with $v_{r} \sim 135 kms^{-1}$, has a very low likelihood to belong to the MW. This star has slightly over-solar iron ([Fe/H]=0.15), and shows underabundant alpha elements ([$\alpha$/Fe]=-0.27) and strongly overabundant n-capture elements (e.g. [CeII/Fe]=0.75). All these seem to be signatures of an extragalactic formation. At the same time, it shows a significant Cu overabundance ([Cu/Fe]=0.25), highly unusual for a disk star. One of the giants appears to be almost surely a MW star (no significant departures from thick-disk abundances), while the other shows, to a lesser extent, the same signatures found in the subgiant ([Fe/H]=0, [$\alpha$/Fe]=-0.15, [Cu/Fe]=0.23, [LaII/Fe]=0.31, [CeII/Fe]=0.21). From these findings, we are lead to believe that a population of extragalactic origin actually exist inside the Canis Mayor overdensity, although the sample is too limited to allow us to push further the interpretation.


%

\end{document}